\documentclass[aps,prb,twocolumn,amsmath,amssymb,nofootinbib,superscriptaddress,floatfix,eqsecnum]{revtex4}

\usepackage{amsmath}
\usepackage{amssymb}
\usepackage{amsthm}
\usepackage[pdftex]{color}
\usepackage{graphicx}
\usepackage{dcolumn} 
\usepackage{bm} 
\usepackage{epic}
\usepackage{longtable}
\usepackage{ulem}   
\newcommand{\bs}[1]{{\boldsymbol{#1}}}

\begin{document}

\title{
Enhancing the stability of a
fractional Chern insulator
against competing phases
      }

\author{Adolfo G.\ Grushin} 
\affiliation{
Instituto de Ciencia de Materiales de Madrid, 
CSIC, Cantoblanco, E-28049 Madrid, 
Spain
            } 
\affiliation{
Physics Department, 
Boston University, 
Boston, Massachusetts 02215, USA
            } 

\author{Titus Neupert} 
\affiliation{
Condensed Matter Theory Group, 
Paul Scherrer Institute, CH-5232 Villigen PSI,
Switzerland
            } 

\author{Claudio Chamon} 
\affiliation{
Physics Department, 
Boston University, 
Boston, Massachusetts 02215, USA
            } 

\author{Christopher Mudry} 
\affiliation{
Condensed Matter Theory Group, 
Paul Scherrer Institute, CH-5232 Villigen PSI,
Switzerland
            } 

\date{\today}

\begin{abstract}
  We construct a two-band lattice model whose bands can carry the
  Chern numbers $C=0,\pm1,\pm2$.
  By means of numerical exact diagonalization, we show that the most
  favorable situation that selects fractional Chern insulators (FCIs)
  is not necessarily the one that mimics Landau levels, namely a flat
  band with Chern number 1.  
  First, we find that the gap, measured in units of the on-site
  electron-electron repulsion, can increase by
  almost two orders of magnitude when the bands
  are flat and carry a Chern number $C=2$ instead of $C=1$. 
  Second, we show that giving a width to the bands
  can help to stabilize a FCI. Finally, we put forward a
  tool to characterize the real-space density profile of the ground
  state that is useful to distinguish FCI from other competing phases
  of matter supporting charge density waves or phase separation.
\end{abstract}

\maketitle


\medskip
\section{Introduction}

A large effort in condensed-matter physics is dedicated to the
classification of distinct phases of matter, exploring their
properties, and establishing their robustness. Many phases of matter
can be classified by the pattern in which a ground-state manifold
breaks spontaneously \textit{and} locally a symmetry of the underlying
many-body Hamiltonian. The experimental discovery in Ref.%
~\onlinecite{Tsui82}
of the fractional quantum Hall effect (FQHE) in high-quality
GaAs-Al${}_{x}$Ga${}_{1-x}$As heterostructures subjected to very large
magnetic field in 1982 was revolutionary in that the FQHE is
associated to a family of phases of matter that are not characterized
by some pattern of spontaneous symmetry breaking with a local order
parameter, but instead by the notion of topological order.%
~\cite{FQHE,Wen-book}
By analogy with spontaneous symmetry breaking and its connection to
(Lie) group theory, a classification of topological order for
many-body Hamiltonians has been undertaken.%
~\cite{Wen-book}

The study of all possible patterns of spontaneous symmetry breaking or
topological order, although immensely useful, fails to address a very
important question.  What is the mechanism by which a given pattern
associated to a phase is established?  This is a quantitative question
that requires deciding in the simplest case of two competing terms in
a many-body Hamiltonian (free energy) which one
is the dominant one. Answering this question, although essential from
a practical point of view, can be very difficult. 
Detailed prescriptions
of the Hamiltonian matter so that nonuniversality rules when
calculating the location in the phase diagram of the boundaries
separating phases of matter. In practice, a realistic answer to this
question requires modern computing power if at all possible.

A problem related to the stability of phases is as follows. Suppose
that we are given two physically different realizations of the same
universal phase: Are the most favorable conditions for realizing the
phase in one physical setting the same as in the other? More
specifically, this paper analyzes the optimal situation for
stabilizing a fractional Chern insulator (FCI). It also
compares and contrasts it to the optimal situation 
for stabilizing a FQHE in a Landau level.

Fractional Chern insulators are strongly correlated phases of
matter for interacting fermions and bosons that have been found in
Bloch bands with vanishing or small bandwidth and nonvanishing Chern
number.%
~\cite{Neupert11a,Sheng11,Wang11a,Regnault11a,Bernevig11,Wang12a,Bernevig12a,Bernevig12b,Wang12b,Liu12a,Laeuchli12,Liu12b,Bernevig-private} 
Fermionic FCIs share the universal properties of the
FQHE that occurs upon partial filling
of Landau levels. It was thus not surprising
that the efforts to establish that FCIs 
can be realized in model Hamiltonians
began by mimicking as closely as
possible the conditions for obtaining the FQHE in Landau levels. 
Bloch bands were chosen so as to try to mirror the properties of 
the Landau levels, which are special because
(I) they support a topological attribute called a Chern number that
equals 1, and
(II) they are independent of their momentum quantum number, i.e., 
they can be thought of as flat bands.
Bloch bands can be made to satisfy properties (I) and (II).%
\cite{Neupert11a}
However,
even if the Bloch bands are made to satisfy conditions (I) and (II), 
it is not \textit{a priori} obvious that energetics allow 
the formation of a fractional quantum Hall ground state.%
~\cite{Parameswaran11,Shankar11,Neupert12a,Estienne12,Murthy12}
For example, the filling fraction 
1/3 is commensurate to the lattice; hence, 
one might expect that,
instead of a fractional Chern state,
interactions select a charge density wave
as the ground state.
Nevertheless, it has become apparent from
exact diagonalization studies of FCIs that the universal properties
of the FQHE can be stabilized by
starting from filling partially a (sufficiently) flat band
of a Chern band insulator for some short-range interaction.

That one can deviate slightly from condition (II) without destroying
the FCI phase is expected, for the topological phase is gapped and
therefore stable to small perturbations. In contrast, one cannot
deviate continuously from condition (I), for the Chern number is an
integer. The work presented below aims to analyze cases when departing
from the conditions (I) and (II), associated to the FQHE in Landau
bands, can lead to enhanced stability of the FCIs.

We present a lattice model for which the Chern
number can take values $C=0,\pm 1,\pm 2$. We how that it is
more difficult to stabilize a FCI with $C=1$ than a FCI with
$C=2$ due to energetics for this model. 
We find that the gaps for a FCI at the
filling fraction $\nu=1/5$ when $C=2$ 
are almost two orders of magnitude larger than
those for the most stable fractional Chern state that we found for
the filling fraction $\nu=1/3$ when $C=1$.
 
To establish that the state at $\nu=1/5$ is a FCI and
not some competing state that breaks spontaneously
and locally some space-group symmetry of the lattice,
we present in this paper a
useful tool to distinguish fractional Chern states
from charge density wave (CDW) states
even when the lattices studied are rather small. We employ this
diagnostic to map out the phase diagram that should emerge in 
the thermodynamic limit as a function of the strength of 
interactions, and to identify the nature of the phases
that compete with the $\nu=1/5$ FCI at $C=2$. 
In doing so, we substantiate our claim that
a FCI can be stabilized by moving away
from property (I) obeyed by the Landau levels.

We then proceed to study the effects of relaxing the flatness
condition (II). We argue that relaxing condition (II)
can help stabilize a FCI. We do this in two ways.

First, we appeal to a general argument that relies on a 
particle-hole transformation of any Hamiltonian $H$
for itinerant and interacting electrons 
with a topological flat band.
By construction, if $H$ supports as a ground state a FCI at 
the filling fraction $\nu$,
then the particle-hole transformed Hamiltonian $\widetilde{H}$
supports a FCI at the filling fraction $\widetilde{\nu}=1-\nu$.
We show that if the interaction is a two-body one 
with translational symmetry, then $\widetilde{H}$ acquires
a one-body contribution through normal ordering. 
This simple observation has the following remarkable 
consequence. A FCI at the filling fraction
$\widetilde{\nu}$ might exist even though $\widetilde{H}$
can be decomposed into a one-body term that generates 
a band-width of the same order as the normal-ordered two-body interaction. 
Conversely,
if this one-body term is switched off adiabatically, the 
ground state at the filling fraction $\widetilde{\nu}$
might undergo a phase transition to a phase that does
not support a topological order. In fact, 
this is the explanation
for the numerical observation that distinct $H$ 
at the filling fractions 
$\nu=1/3$ or $\nu=1/5$ support FCIs as the ground states,
while the very same $H$ at the filling fractions 
$\widetilde{\nu}=2/3$
or $\widetilde{\nu}=4/5$ are not topologically ordered.%
~\cite{Footnote: resolution discrepency with Bernevig,Neupert11b,Bernevig-KITPC}

Second,
in order to probe the phase diagram of the system as
a function of two parameters, 
one of which controls the size of the
bandwidth, we use another numerical tool 
to help trace out the phase
boundaries of topological states using 
exact diagonalization studies
of small lattices. We look at the ratio between 
the observed gap
$\Delta$ and the spread in energy $\delta$ of 
those states that belong to
the manifold of states that are to become degenerate in the
thermodynamic limit. If $L$ is the characteristic linear size 
of the lattice, this ratio
$
\Delta/\delta\sim \exp(L/\xi)
$
entails information on the correlation length $\xi$ 
of the system. The correlation length $\xi$ 
is a useful mean to measure the distance to the phase boundary.
By combining the parametric dependence of $\Delta$ along
contours at constant $\Delta/\delta$,
we can estimate the region of stability of the FCI at $\nu=1/5$
and $C=2$ from exact diagonalization. 
We find that moving away from condition (II) that is obeyed by
Landau levels, i.e., including a non-zero bandwidth, 
can (up to a limit) help increase 
(and not decrease) the range of phase stability.

\section{ 
The model
        }
\label{sec: Model}

\begin{figure}
\includegraphics[angle=0,scale=0.4]{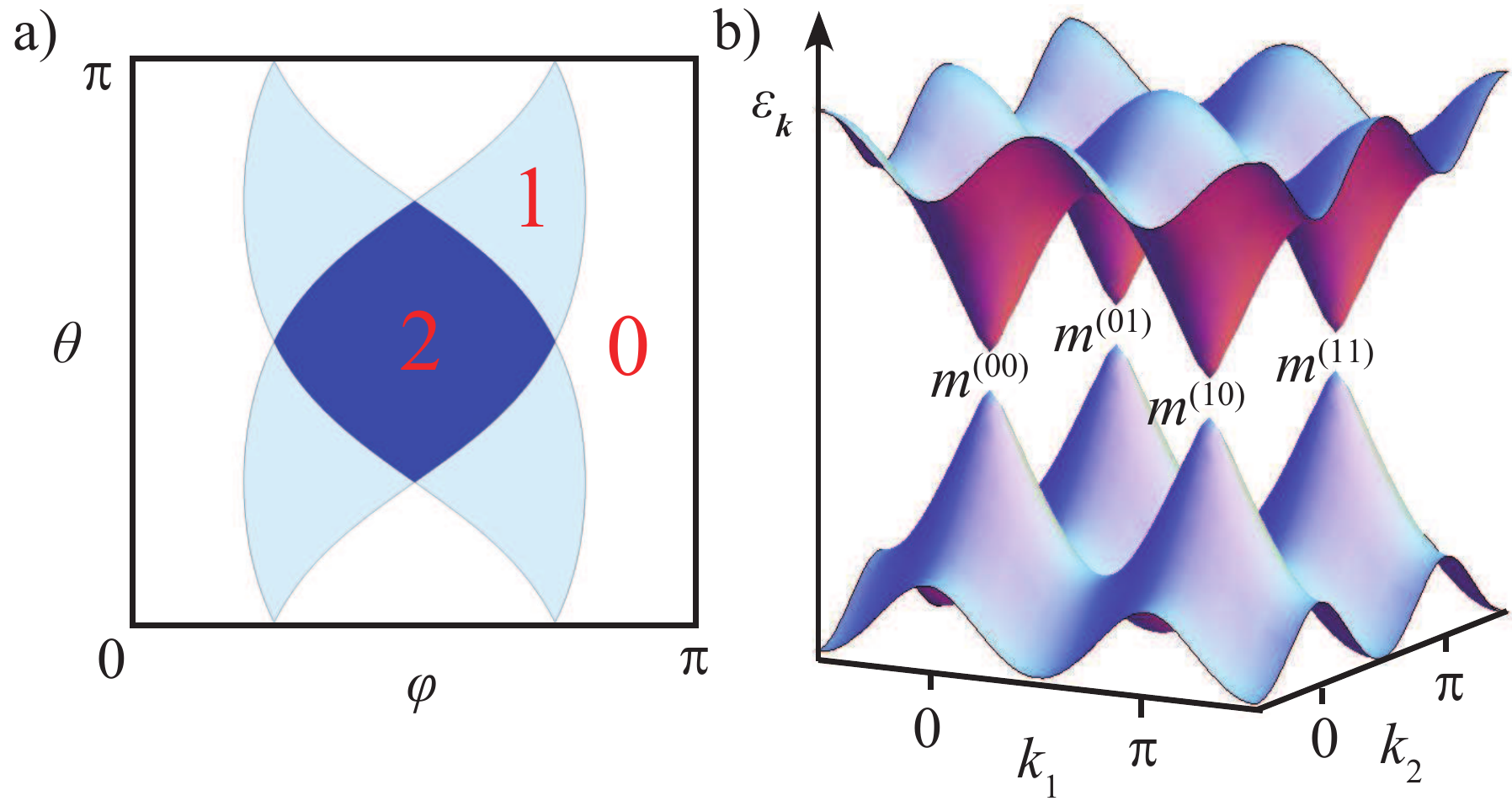}
\caption{\label{fig: single-particle band structure}
(Color online)
(a) Any one of the two Chern numbers 
of the two bands of the noninteracting Hamiltonian%
~\eqref{eq: noninteracting H momentum space} 
depends on the direction of the four-component mass vector 
$h=(h^{\,}_\mu)$. When restricted to the form 
$
h=
|h|
\Big(\cos\varphi,\sin\varphi\,\cos\theta,0,(1/2)\sin\varphi\sin\theta\Big)
$, 
the Chern number phase diagram is obtained as shown 
(white, $|C|=0$; light, $|C|=1$; dark, $|C|=2$).
(b) Band structure of the noninteracting Hamiltonian%
~\eqref{eq: noninteracting H momentum space} 
with mass terms $m^{(ij)},\ i,j=0,1$, 
defined in Eq.~\eqref{eq: massterms}.
        }
\end{figure}

Consider a noninteracting tight-binding model for fermions on 
the two-dimensional square lattice $\Lambda$ 
made of $L^{\,}_{1}\times L^{\,}_{2}$ 
sites and spanned by the orthonormal primitive lattice vectors 
$\hat{\bs{e}}^{\,}_{1}$ and $\hat{\bs{e}}^{\,}_{2}$. 
We impose periodic boundary conditions.
The fermions have two internal degrees of freedom per site, 
which we denote as a spin degree of freedom 
$s=\uparrow,\downarrow$. 
Each hopping process is associated with a spin flip. 
The resulting noninteracting Bloch Hamiltonian 
supports two bands and reads in momentum space
\begin{subequations}
\begin{eqnarray}
&&
H^{\,}_{0}:=
\sum_{\bs{k}\in\mathrm{BZ}}
c^{\dagger}_{\bs{k}}
\,\bs{B}^{\,}_{\bs{k}}\cdot\bs{\sigma}\,
c^{\,}_{\bs{k}},
\\
&&
B^{\,}_{\bs{k};1}+\mathrm{i}\,B^{\,}_{\bs{k};2}:=
t(\sin\,k^{\,}_{1}+\mathrm{i}\,\sin \, k^{\,}_{2}),
\\
&&
B^{\,}_{\bs{k};3}:=
h^{\,}_{1}\cos\,k^{\,}_{1}
+h^{\,}_{2}\cos \, k^{\,}_{2}
+h^{\,}_{3}
\nonumber
\\
&&
\hphantom{B^{\,}_{\bs{k};3}:=}
+
h^{\,}_{4}
\left[
\cos(k^{\,}_{1}+k^{\,}_{2})+\cos(k^{\,}_{1}-k^{\,}_{2})
\right],
\nonumber\\
&&
\end{eqnarray}
where 
$
c^{\dag}_{\bs{k}}\equiv
(c^{\dag}_{\bs{k},\uparrow},c^{\dag}_{\bs{k},\downarrow})
$ 
and 
$c^{\dag}_{\bs{k},s}$ 
creates a fermion at momentum $\bs{k}$ in the Brillouin zone 
(BZ) 
with spin $s=\uparrow, \downarrow$
while $\bs{\sigma}=(\sigma^{\,}_{1},\sigma^{\,}_{2},\sigma^{\,}_{3})$ 
are the three Pauli matrices acting on spin space.
The parameters $t$ and $h^{\,}_\mu,\ \mu=1,\cdots,4$, are real.

The role of the hopping parameters $t$ and 
$h^{\,}_\mu,\ \mu=1,\cdots,4$, 
can be illustrated by fixing $t>0$ 
and expanding the Hamiltonian around 
the four inversion-symmetric momenta in the BZ 
$\bs{k}^{(ij)}=\pi(i,j),\ i,j=0,1$.
To linear order in the deviation 
$\bs{p}^{(ij)}=\bs{k}-\bs{k}^{(ij)}$, $i,j=0,1$, 
from each of these four momenta, 
the Hamiltonian takes a Dirac form with masses given by
\label{eq: noninteracting H momentum space}
\end{subequations}
\begin{subequations}
\begin{equation}
m^{(ij)}=
(-1)^i h^{\,}_{1}
+(-1)^j h^{\,}_{2}
+h^{\,}_{3}
+(-1)^{i+j}\,2\,h^{\,}_{4}.
\label{eq: massterms}
\end{equation}
The model thus makes it possible to independently control the sign and 
magnitude of the mass at each of the four Dirac points via 
the parameters $h^{\,}_{\mu},\ \mu=1,\cdots,4$. 
If all the Dirac points have a nonvanishing gap, 
the Chern number of each of the two bands is well defined. 
Each Dirac point contributes $\pm1/2$ to the Chern number, 
with the sign depending on the chirality 
$\exp(\mathrm{i}\,k^{(ij)}_{1}+\mathrm{i}\,k^{(ij)}_{2})$
of the kinetic part of the Dirac Hamiltonian 
and the sign of the mass gap. 
The total Chern number of the lower band is then given by
\begin{equation}\label{eq: Cnumber}
C=\frac{1}{2}\sum_{i,j=0,1} (-1)^{i+j}\,\mathrm{sgn}\,m^{(ij)}
\end{equation}
\end{subequations}
and therefore can assume the values $C=\pm2,\pm1,0$ 
in our two-band model 
(see Fig.~\ref{fig: single-particle band structure}).

In what follows, we study this model in presence of 
density-density interactions between fermions 
on the same lattice site ($U$) 
and between fermions on neighboring sites ($V$) 
\begin{equation}
H^{\,}_{\mathrm{int}}:=
\frac{U}{2}
\sum_{\bs{r}\in\Lambda}
\sum_{s\neq s'}
\rho^{\,}_{\bs{r},s }\,
\rho^{\,}_{\bs{r},s'}
+
\frac{V}{2}
\sum_{(\bs{r},\bs{r}')}
\sum_{s,s'}
\rho^{\,}_{\bs{r} ,s }\,
\rho^{\,}_{\bs{r}',s'}
\label{eq: def H int}
\end{equation}
upon partial filling the lower band.
Here, $(\cdot,\cdot)$ 
denotes nearest-neighbor lattice sites and 
$
\rho^{\,}_{\bs{r},s}:=
c^{\dag}_{\bs{r},s}\,
c^{\,}_{\bs{r},s}
$ 
is the density of fermions with spin 
$s=\uparrow,\downarrow$ at site $\bs{r}\in\Lambda$. 

We are interested in the situation where 
the following hierarchy of energy scales applies
\begin{subequations}
\begin{equation}
W\ll U,V\ll m,
\label{eq: hierarchy energies}
\end{equation}
with $W$ and $m$ being the bandwidth and the band gap 
of the noninteracting band structure, respectively.

The fact that the band gap $m$ is the largest energy scale 
justifies considering the ideal limit
\begin{equation}
W/m\ll U/m,V/m\to0,
\end{equation}
\end{subequations}
in which
the single-particle Hilbert space is projected onto
the subspace spanned by the states in the lower band.
Hence, we project the interaction%
~(\ref{eq: def H int})
onto the Fock space built out of the single-particle 
subspace of the lower band. This projection gives
\begin{subequations}
\label{eq: H that is diagonalized}
\begin{eqnarray}
&&
H^{\mathrm{pro}}_{\mathrm{int}}:=
\frac{-1}{L^{\,}_{1}L^{\,}_{2}}
\sum_{\bs{k},\bs{k}',\bs{q}}
\gamma^{\,}_{\bs{k},\bs{k}',\bs{q}}\,
\chi^{\dag}_{\bs{k}+\bs{q}}\,
\chi^{\dag}_{\bs{k}'-\bs{q}}\,
\chi^{\,}_{\bs{k} }\,
\chi^{\,}_{\bs{k}'},
\\
&&
\gamma^{\,}_{\bs{k},\bs{k}',\bs{q}}:=
\sum_{s,s'}
\left[
\frac{U}{2}(1-\delta^{\,}_{s,s'})
+
V(\cos\, q^{\,}_{1}+\cos\, q^{\,}_{2})
\right],
\nonumber\\
&&
\qquad\qquad\
\times
u^{\,}_{\bs{k}+\bs{q},s}
u^{\,}_{\bs{k}'-\bs{q},s'}
u^{*}_{\bs{k},s}
u^{*}_{\bs{k}',s'},
\end{eqnarray}
\end{subequations}
where $u^{\,}_{\bs{k}}\equiv(u^{\,}_{\bs{k},s})$ 
is the eigenvector of the lower band of the $2\times2$ Bloch 
Hamiltonian $\bs{B}^{\,}_{\bs{k}}\cdot\bs{\sigma}$ in Eq.%
~\eqref{eq: noninteracting H momentum space}
at momentum $\bs{k}$, while $\chi^{\dag}_{\bs{k}}$ 
is the second quantized operator that creates 
the corresponding state in the lower band.

The fact that $W$ is the smallest energy scale
justifies considering the ideal limit in which the lower band is flat.
This is achieved by defining the kinetic energy 
(see Ref.~\onlinecite{Neupert11a}),
\begin{equation}
H^{\mathrm{flat}}_{0}:=
\sum_{\bs{k}\in\mathrm{BZ}}
c^{\dag}_{\bs{k}}
\frac{
\bs{B}^{\,}_{\bs{k}}\cdot\bs{\sigma}
     }
     {
|\bs{B}^{\,}_{\bs{k}}|
     }
c^{\,}_{\bs{k}}.
\label{eq: noninteracting H flat}
\end{equation}
Here, dividing $\bs{B}^{\,}_{\bs{k}}\cdot\bs{\sigma}$ by 
$|\bs{B}^{\,}_{\bs{k}}|$ amounts to assigning to all Bloch states
from the lower band the energy $-1$ 
and to all Bloch states from the upper band the energy $+1$. 
This deformation of the Hamiltonian induces 
arbitrary-range hopping amplitudes in position space, 
but preserves its locality in the sense that 
these hopping amplitudes decay exponentially with distance.%
~\cite{Neupert11a} 
Interpolating between $H^{\,}_{0}$ and $H^{\mathrm{flat}}_{0}$ 
through the parametric dependence on $0\leq\lambda\leq1$ of
\begin{equation}
H^{\,}_{0}(\lambda):=
(1-\lambda)H^{\mathrm{flat}}_{0}
+
\lambda H^{\,}_{0}
\label{eq: def H0 lambda}
\end{equation}
makes it possible to choose the bandwidth at will. 

We begin with the study of the case $\lambda=0$, using the
Hamiltonian 
\begin{equation}
H(\lambda):=
H^{\,}_{0}(\lambda)
+
H^{\mathrm{pro}}_{\mathrm{int}}
\label{eq: def Hamiltonian studied by exact dia}
\end{equation} 
by carrying out exact diagonalization studies of small systems 
in Secs.~\ref{sec: C1 and C2} and \ref{sec: CDWpd}. In Sec.%
~\ref{sec:  dispersion} 
we study how a finite bandwidth 
(obtained by varying $\lambda$) 
affects the fractional Chern states.

\section{
$C=1$ and $C=2$ 
fractional Chern states
        }
\label{sec: C1 and C2}

\begin{figure}
\hskip -30 true pt
\includegraphics[angle=0,scale=0.55]{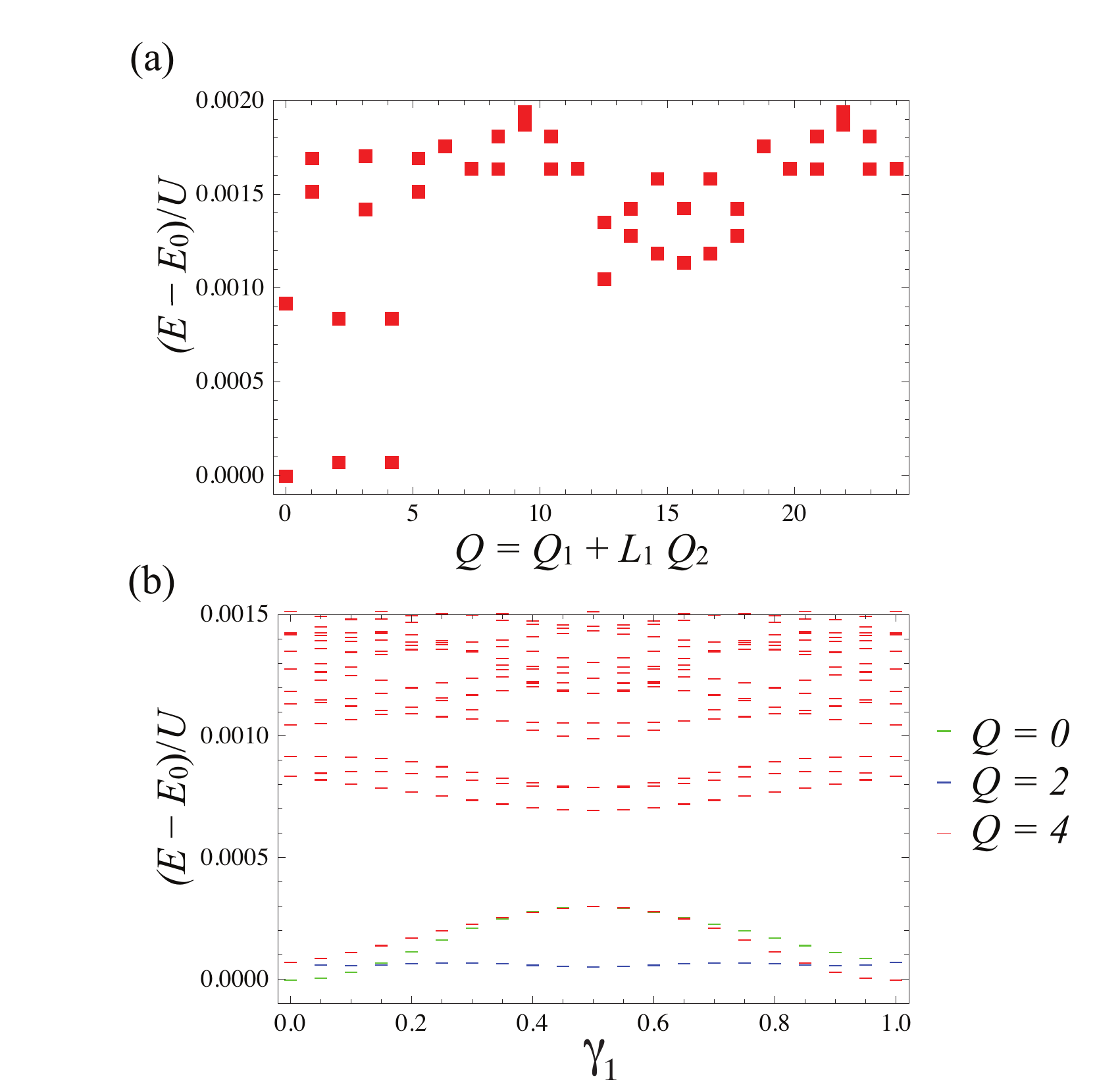}
\caption{\label{fig: thirdstatespectraflux}
(Color online)
(a) Energy eigenvalues of Hamiltonian
(\ref{eq: def Hamiltonian studied by exact dia})
at $\lambda=0$
measured relative to the ground-state energy
at the filling fraction 
$\nu=1/3$ for $L=6\times4$ and $N=8$ particles
as a function of the total momentum $\bs{Q}$. 
The noninteracting parameters 
of Hamiltonian~(\ref{eq: def Hamiltonian studied by exact dia})
are 
$h^{\,}_{4}=0$ and $h^{\,}_{1}=h^{\,}_{2}=-h^{\,}_{3}=-t$, 
corresponding to a noninteracting band-with $C=1$.
The interacting parameters 
of Hamiltonian~(\ref{eq: def Hamiltonian studied by exact dia})
are $U=t$ and $V=0$.
A threefold quasidegenerate ground state is observed 
in agreement with Ref~\onlinecite{Bernevig12a}. 
(b) Spectral flow induced by a flux insertion 
in the $\gamma^{\,}_{1}$ direction for $N=8$
particles. The three lowest-lying states do not mix with the 
would-be continuum of excited states and return 
to the same configuration 
after three flux quanta have been inserted 
(only the insertion of one flux quantum is shown). 
For comparison with Fig.~(\ref{fig: fifthstatespectraflux}),
the FCI at $\nu=1/3$ has disappeared when $V=t$.
        }
\end{figure}

The topological properties of the anticipated FCIs
are believed to be described by a hierarchical 
Chern-Simons theory%
~\cite{Read90,Blok90,Blok91,Wen92a,Wen92b}
with the action
\begin{equation}
S^{\,}_{\mathrm{CS}}:=
\frac{\varepsilon^{\,}_{\mu\nu\lambda}}{4\pi}
\int\mathrm{d}^{2}\bs{r}\,\mathrm{d} t\,
\left(
-K^{\,}_{ij}\, a^{\mu}_{i}\partial^{\nu} a^{\lambda}_{j}
+
2e\,
Q^{\,}_{i} a^{\mu}_{i}\partial^{\nu} A^{\lambda}
\right)
\label{eq: CS action}
\end{equation}
in the continuum limit. Here
$a^{1}_i$, $a^{2}_i$, and $a^{0}_i$ 
are the two spatial components and the temporal component of 
$i=1,\cdots,N^{\,}_{\mathrm{f}}$ 
flavors of hydrodynamical gauge fields.
The electromagnetic gauge field is denoted by
$A^{\mu},\ \mu=0,1,2$.
The $N^{\,}_{\mathrm{f}}\times N^{\,}_{\mathrm{f}}$ matrix
$K$ is symmetric with integer entries.
The $N^{\,}_{\mathrm{f}}$-component vector
$Q=(Q^{\,}_{i})$ is the charge vector and it has integer entries. 
The summation convention over the repeated flavor
and Greek indices is implied. 

The topological ground-state degeneracy on 
the torus and the Hall conductivity are given by 
$\mathrm{det}\,K$ and 
\begin{equation}
\sigma^{\,}_{\mathrm{H}}=
\frac{e^2}{h}\, Q^{\,}_{i}\,K^{-1}_{ij}\, Q^{\,}_j,
\end{equation}
respectively.
On the other hand, we may naively expect the Hall
conductivity of the fractional Chern states to be
\begin{equation}
\sigma^{\,}_{\mathrm{H}}=\frac{e^2}h C\,\nu,
\label{eq:sigma-Hall}
\end{equation}
where $\nu$ is the filling fraction of the lower band. 
When the band is fully filled ($\nu=1$), 
indeed the expectation is fulfilled. That this formula holds 
for generic filling is always \textit{permitted} 
but it is \textit{not guaranteed}.
When the Berry curvature is not constant in the BZ,
$\sigma^{\,}_{\mathrm{H}}$ does not need to be tied to 
$\nu$ anymore.%
~\cite{Kol93,Neupert-sigma-paper}

It is known that topologically degenerate ground states 
on the torus can be transferred into one another by 
adiabatically inserting flux 
through the torus.~\cite{Thouless89}
After an insertion of an integer number of flux quanta equal 
to the degeneracy, one must recover the same state. 
For a many-body state $|\Psi\rangle$ with $N$ 
particles inserting a flux $2\pi \gamma^{\,}_{i}$ 
in the $\hat{\bs{e}}^{\,}_{i}$ direction
is equivalent to imposing the twisted boundary conditions
\begin{equation}
\begin{split} 
&
\langle
\bs{r}^{\,}_{1},\cdots,\bs{r}^{\,}_{j}
+
L^{\,}_{i}\hat{\bs{e}}_{i},\cdots,\bs{r}^{\,}_N|
\Psi\rangle
\\
&
\quad
=
e^{
\mathrm{i}\,2\pi\,\gamma^{\,}_{i}
  }
\langle
\bs{r}^{\,}_{1},\cdots,
\bs{r}^{\,}_{j},\cdots,\bs{r}^{\,}_{N}|\Psi
\rangle,
\qquad j=1,\cdots,N.
\end{split}
\label{eq: twisted boundary conditions}
\end{equation}
        
With this background to 
FCIs in mind, 
we proceed to discuss model%
~\eqref{eq: noninteracting H momentum space}. 
The principal advantage of this model is that it makes it possible 
to change the Chern number of the band from $C=-2$ to 2, 
in steps of 1. We first focus on the case $C=1$, 
which can be realized by choosing
$h^{\,}_{1}=h^{\,}_{2}=-h^{\,}_{3}=-t$ and $h^{\,}_{4}=0$ 
and was previously studied in Ref.~\onlinecite{Bernevig12a}. 
In this case, the model hosts a $1/3$ 
fractional Chern insulator state 
which we here include for completeness.

Figure \ref{fig: thirdstatespectraflux}(a) 
shows the low-energy portion of the many-body spectrum and 
the expected threefold quasidegenerate 
fractional Chern ground state
when $U=t$ and $V=0$. 
The spectrum is plotted as a function of the total 
center-of-mass momentum $\bs{Q}$. 
This is a good quantum number because the many-body Hamiltonian
shares the space-group symmetry of the square lattice. 
The fact that the states fall in the sectors 
with center-of-mass momenta 
$
(Q^{\,}_{1},Q^{\,}_{2})_{6\times 4}=
\left\lbrace(0,0), (2,0), (4,0)\right\rbrace
$ 
agrees with the counting rule introduced in 
Ref.~\onlinecite{Regnault11a}. 
The threefold quasidegeneracy is consistent 
with a Chern-Simons theory%
~\eqref{eq: CS action} 
for a single species of gauge fields and $K=3$, $Q=1$.

Figure \ref{fig: thirdstatespectraflux}(b) 
shows the evolution of the spectrum when flux is 
adiabatically inserted 
in the $\hat{\bs{e}}^{\,}_{1}$ direction for a $6\times 4$ 
site lattice with $U=t$, $V=0$. 
One observes that the quasidegenerate ground-state 
manifold indeed evolves independently from the excited states. 
This observation is again consistent with a 
$\nu=1/3$ fractional Chern state. Moreover, 
we checked that this state survives when $h^{\,}_{4}\neq 0$ 
as long as the Chern number remains $C=1$.

With the help of Eq.~\eqref{eq: Cnumber} 
one verifies that for sufficiently large $h^{\,}_{4}>0$, 
the Chern number becomes $C=2$. 
In fact, the simplest set of parameters
that hosts a Bloch band with $C=2$ is 
$h^{\,}_{1}=h^{\,}_{2}=h^{\,}_{3}=0$ 
and $h^{\,}_{4}\neq 0$. 
For this set of the parameters, model%
~\eqref{eq: noninteracting H momentum space} 
can be reinterpreted as a layered model 
where two square lattices are 
superimposed on each other without any hopping term 
that connects them. 
We are going to argue that when
$h^{\,}_{1}=h^{\,}_{2}=h^{\,}_{3}=0$,
$h^{\,}_{4}=0.7\,t$, and $U=V=t$,
the many-body Hamiltonian
$H(\lambda=0)$ 
defined in Eq.~(\ref{eq: def Hamiltonian studied by exact dia})
realizes a FCI at $\nu=1/5$ filling.

\begin{figure}
\includegraphics[angle=0,scale=1.3]{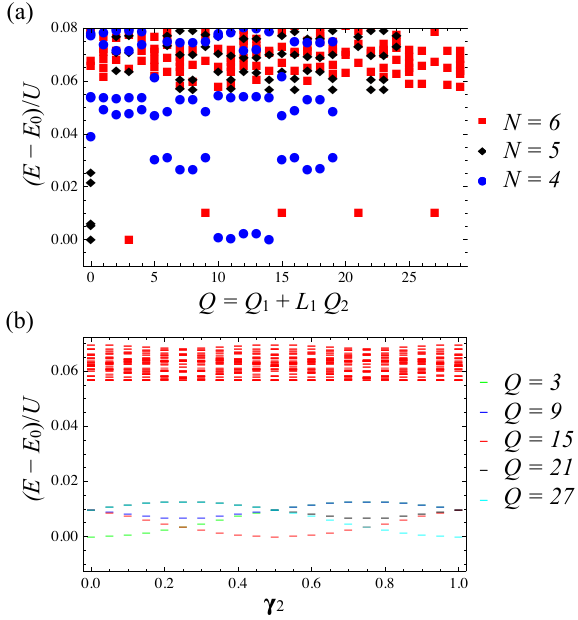}
\caption{
(Color online)
(a) Energy eigenvalues 
of Hamiltonian~(\ref{eq: def Hamiltonian studied by exact dia})
at $\lambda=0$
measured relative to the ground state energy
at the filling fraction $\nu=1/5$ 
for three different system sizes corresponding to 
$N=6,5$, and $4$ particles 
(green diamonds, red squares, and black circles, respectively).
The noninteracting parameters 
of Hamiltonian~(\ref{eq: def Hamiltonian studied by exact dia})
are
$h^{\,}_{4}=0.7\,t$ and $h^{\,}_{i}=0$ for $i=1,2,3$ 
corresponding to $C=2$.
The interacting parameters 
of Hamiltonian~(\ref{eq: def Hamiltonian studied by exact dia})
are $U=V=t$. All three systems show 
a fivefold quasidegenerate ground state (see text for details). 
(b) Spectral flow induced by a flux insertion 
in the $\gamma^{\,}_{2}$ direction for $N=6$
particles. The five lowest lying states do not mix with the 
would-be continuum of excited states and return 
to the same configuration 
after five flux quanta have been inserted 
(only the insertion of one flux quantum is shown). 
        }
\label{fig: fifthstatespectraflux}
\end{figure}

First, we show in 
Fig.~\ref{fig: fifthstatespectraflux}(a) that for system sizes 
$6\times 5$, $5\times 5$, and $5\times 4$ with 
$N=L^{\,}_{1}L^{\,}_{2}/5$ particles, 
the many-body eigenvalues possess a fivefold degenerate 
ground state. 
The total momentum sectors $\bs{Q}$ at which these states 
should appear 
for a fractional Chern state can be calculated with 
a generalization 
of the counting rule of Ref.~\onlinecite{Regnault11a}. 
For a $L^{\,}_{1}\times L^{\,}_{2}$ lattice with $N$ particles 
where we choose $L^{\,}_{1}$ 
to be a multiple of five, the momentum sectors are given by 
\begin{subequations}
\begin{equation}
\begin{pmatrix}
Q^{\,}_{1}
\\
Q^{\,}_{2}
\end{pmatrix}=
\begin{pmatrix}
[N(L^{\,}_{1}-5)/2+m N]\ \mathrm{mod}\, L^{\,}_{1}
\\
N(L^{\,}_{2}-1)/2\ \mathrm{mod}\, L^{\,}_{2}
\end{pmatrix},
\end{equation}
with $m=0,1,2,3,4$. 
If $L^{\,}_{2}$ is a multiple of five the same rule applies 
upon interchanging $Q^{\,}_{1}$ for $Q^{\,}_{2}$. 
For the square lattices made of
$6\times 5$, $5\times 5$, and $5\times 4$
sites, the momentum sectors should fall at 
\begin{equation}
\begin{split}
&
\begin{pmatrix}
Q^{\,}_{1}
\\
Q^{\,}_{2}
\end{pmatrix}^{\,}_{6\times 5}=
\begin{pmatrix}
3
\\
0
\end{pmatrix},
\begin{pmatrix}
3
\\
1
\end{pmatrix},
\begin{pmatrix}
3
\\
2
\end{pmatrix},
\begin{pmatrix}
3
\\
3
\end{pmatrix},
\begin{pmatrix}
3
\\
4
\end{pmatrix},
\\
&
\begin{pmatrix}
Q^{\,}_{1}
\\
Q^{\,}_{2}
\end{pmatrix}^{\,}_{5\times 5}=
\begin{pmatrix}
0
\\
0
\end{pmatrix},
\begin{pmatrix}
0
\\
0
\end{pmatrix},
\begin{pmatrix}
0
\\
0
\end{pmatrix},
\begin{pmatrix}
0
\\
0
\end{pmatrix},
\begin{pmatrix}
0
\\
0
\end{pmatrix},
\\
&
\begin{pmatrix}
Q^{\,}_{1}
\\
Q^{\,}_{2}
\end{pmatrix}^{\,}_{5\times 4}=
\begin{pmatrix}
0
\\
2
\end{pmatrix},
\begin{pmatrix}
1
\\
2
\end{pmatrix},
\begin{pmatrix}
2
\\
2
\end{pmatrix},
\begin{pmatrix}
3
\\
2
\end{pmatrix},
\begin{pmatrix}
4
\\
2
\end{pmatrix},
\end{split}
\end{equation}
\end{subequations}
respectively. This is exactly what we find 
[see Fig.~\ref{fig: fifthstatespectraflux}(a)].  

Second, 
Fig.~\ref{fig: fifthstatespectraflux}(b) 
shows the flux insertion in the $\hat{\bs{e}}^{\,}_{2}$ 
direction for lattice size $L=6\times 5$. 
The five states corresponding to 
the ground-state manifold evolve independently 
from the continuum 
and they only recover their original position after 
five flux quanta 
have been inserted in the system. 

Third,
we compute the Hall conductivity of the five quasidegenerate 
ground states via their many-body Chern number. 
To this end, we apply the discretized version,
\begin{subequations}
\label{eq: tilde sigma H subeqs}
\begin{equation}
\widetilde{\sigma}^{\,}_{\mathrm{H}}:=
\frac{e^2}{h}\,\frac{2\pi}{L^{\,}_{1}L^{\,}_{2}}
\sum_{\bs{k}\in\mathrm{BZ}}
F^{\,}_{\bs{k}}\,
\bar{n}^{\,}_{\bs{k}}
\label{eq: tilde sigma H}
\end{equation}
of the formula derived in Ref.~\onlinecite{Neupert12b}
for the quantum Hall conductivity $\sigma^{\,}_{\mathrm{H}}$
as befits a finite lattice.
Here, the single-particle Berry curvature $F^{\,}_{\bs{k}}$ 
and the many-body occupation number $\bar{n}^{\,}_{\bs{k}}$ 
averaged over the five quasidegenerate 
ground states $|\Psi^{\,}_i\rangle$, $i=1,\cdots,5$,
are given by
\begin{equation}
F^{\,}_{\bs{k}}:=
\mathrm{i}
\sum^{\,}_s
\partial^{\,}_{k^{\,}_{2}} u^{*}_{\bs{k},s} 
\partial^{\,}_{k^{\,}_{1}} 
u^{\,}_{\bs{k},s}
-
(1\leftrightarrow 2)
\label{eq: F of k}
\end{equation}
and
\begin{equation}
\bar{n}^{\,}_{\bs{k}}:= 
\frac{1}{5L^{\,}_{1}L^{\,}_{2}}
\sum_{i=1}^{5}
\left\langle
\Psi^{\,}_{i}
\left|
\chi^{\dag}_{\bs{k}}
\chi^{\,}_{\bs{k}}
\right|
\Psi^{\,}_{i}
\right\rangle,
\label{eq: n of k}
\end{equation}
\end{subequations}
respectively. It is shown in Ref.~\onlinecite{Neupert12b} that
$\widetilde{\sigma}^{\,}_{\mathrm{H}}$
converges to the Hall conductivity 
$\sigma^{\,}_{\mathrm{H}}$
averaged over the degenerate ground states 
in the thermodynamic limit,
provided no spontaneous symmetry-breaking of 
translation invariance occurs.
Observe that the accuracy of the quantization of 
$\widetilde{\sigma}^{\,}_{\mathrm{H}}$
is limited by the finite size of the system, 
as the Berry curvature~(\ref{eq: F of k})
is only summed over 
$L^{\,}_{1}\times L^{\,}_{2}$ 
points in the BZ to replace an integral 
in the thermodynamic limit.
We have evaluated Eq.~(\ref{eq: tilde sigma H subeqs})
for the values
$h^{\,}_{1}=h^{\,}_{2}=h^{\,}_{3}=0$,
$h^{\,}_{4}=0.7\,t$, and $U=V=t$
in the many-body Hamiltonian
$H(\lambda=0)$ 
defined by Eq.~(\ref{eq: def Hamiltonian studied by exact dia})
for the rectangular lattices made of
$5\times5$,
$5\times6$,
$3\times10$,
and
$2\times15$
sites. We obtain 
for $\widetilde{\sigma}^{\,}_{\mathrm{H}}$
in units of $e^2/h$ the values 
0.391,
0.401,
0.400,
and
0.500,
respectively.
Results with the same numerical accuracy are obtained 
for each ground state individually, i.e., 
without the averaging over all ground states in 
Eq.~\eqref{eq: n of k}.
We conclude that Eq.~\eqref{eq:sigma-Hall} 
with $C=2$ and $\nu=1/5$ is captured by
Eq.~(\ref{eq: tilde sigma H}) 
within $2\%$ accuracy for the rectangular lattices
with the aspect ratios 
$5\times5$,
$5\times6$,
$3\times10$.
The value of $\widetilde{\sigma}^{\,}_{\mathrm{H}}$
for the aspect ratio $2\times15$ deviates from
$\sigma^{\,}_{\mathrm{H}}=0.4 e^2/h$ by 25$\%$.
However, a ladder with one leg obeying periodic boundary conditions
is topologically equivalent to a ring, not to a torus as is the case
for the $5\times5$, $5\times6$, $3\times10$ lattices.

All three observations are consistent 
with a FCI 
described by the Chern-Simons theory~\eqref{eq: CS action} 
for two species of gauge fields and 
\begin{equation}
K=
\begin{pmatrix}
3&2
\\
2&3
\end{pmatrix},
\qquad
Q=
\begin{pmatrix}
1
\\
1
\end{pmatrix}.
\end{equation}
Note that a single species of gauge fields with $K=5$ 
would not explain the Hall conductivity 
$\sigma^{\,}_{\mathrm{H}}=2e^2/(5h)$.

Although these three 
pieces of evidence already point towards 
a $1/5$ fractional Chern state, this state could still 
be a CDW. To show that this is not the case,
we put forward in Sec.~\ref{sec: CDWpd}
a different tool to distinguish liquid states such as 
the fractional Chern states from other competing orders, 
namely phases with broken translational invariance such 
as CDW or phase-separated phases. 

\medskip
\section{
Topological order and local symmetry breaking
        }
\label{sec: CDWpd}

Both topological order and long-range order with a local order
parameter are emergent phenomena that occur in 
the thermodynamic
limit for a given dimensionality of space.
Most model Hamiltonians are not exactly solvable in the
thermodynamic limit. Approximations such as
variational methods can be used to access the 
thermodynamic limit,
but they are uncontrolled. Exact diagonalization techniques
are not biased, but they are limited to finite sizes. 
The extrapolation of exact finite-size spectra to the 
thermodynamic limit must be undertaken with great care.
In particular, this extrapolation can involve
subtle dimensional crossovers (e.g.,
the thin torus limit discussed below).

In view of the effects of finite-size corrections to the
thermodynamic limit, it is essential to use complementary
probes for topological order in exact diagonalization studies
to argue convincingly that a candidate FCI
has the featureless character 
that is demanded from a topological fluid and to map out its
boundary in the phase diagram.
The particle entanglement spectrum has been used
in Ref.~\onlinecite{Bernevig12a}
to identify FCIs.
We are going to give another criterion that
distinguishes a fractional Chern phase from a
competing phase that breaks spontaneously
the space-group symmetry of the lattice.
Examples of such competing phases are 
CDW and macroscopic phase separation.
Equipped with this diagnostic, we are going to map
out the phase diagram as a function of 
the dimensionless parameter
$-2<V/U<1$ of Hamiltonian%
~(\ref{eq: def Hamiltonian studied by exact dia})
with a flat band, i.e., $\lambda=0$. 

Let $O^{\,}_{\bs{r}}$ be any local operator defined for
any site $\bs{r}$ from the lattice $\Lambda$.
We assume that the ground-state manifold of 
$H(\lambda=0)$ is $n$-dimensional and 
spanned by the quasidegenerate ground states
$|\Psi^{\,}_{1}\rangle,\cdots,|\Psi^{\,}_{n}\rangle$.
By assumption, $H(\lambda=0)$ shares the space-group symmetry of
the lattice $\Lambda$. Hence, we can always choose
$|\Psi^{\,}_{i}\rangle$ 
to be a simultaneous eigenstate of the momentum operator
with the center of mass $\bs{Q}^{\,}_{i}$
where $i=1,\cdots,n$. We define  the $n\times n$ 
Hermitian matrix ${\cal O}^{\,}_{\bs{r}}$ with elements
\begin{equation}
{\cal O}_{\bs{r};ij}:=
\langle\Psi^{\,}_{i}|\,
O^{\,}_{\bs{r}}\,
|\Psi^{\,}_{j}\rangle,
\end{equation}
which amounts to restricting the operator 
$O^{\,}_{\bs{r}}$
to its action on the ground-state manifold.
Let $v^{(i)}_{O;\bs{r}}$ 
be the $i$-th eigenstate of the matrix
${\cal O}^{\,}_{\bs{r}}$.  
Its eigenvalue $\lambda_O^{(i)}$ is independent
of the lattice site $\bs{r}$ 
as a consequence of translation symmetry of the Hamiltonian.%
~\cite{footenote: kappa(i) does not depend on r} 
If the space-group symmetry of the lattice $\Lambda$ 
and of Hamiltonian $H(\lambda=0)$ is not to be
broken spontaneously by the ground-state manifold, then it is
necessary for all eigenvalues 
$\lambda^{(1)}_{O},\dots,\lambda^{(n)}_{O}$
of ${\cal O}^{\,}_{\bs{r}}$ to be equal. 
The spread of the eigenvalues
measures the degree by which the symmetry associated with the
operator $O^{\,}_{\bs{r}}$ is broken. If 
$\lambda^{\mathrm{min}}_{O}$ 
and $\lambda_O^{\rm max}$ are, respectively, 
the minimum and maximum eigenvalues, 
the quantity to monitor is the difference
\begin{equation}
\delta \lambda_O:=
\lambda_O^{\rm max}
-
\lambda_O^{\rm min}
\;.
\end{equation}
We demand that the condition 
$\delta \lambda_O\to 0$ holds in a
suitable thermodynamic limit for all local operators 
if topological order is to hold.

We now apply this analysis for the case when 
$O^{\,}_{\bs{r}}=\rho^{\,}_{\bs{r}}$,
the local density operator being defined by
\begin{subequations}
\begin{equation}
\rho^{\,}_{\bs{r}}:=
\frac{1}{L^{\,}_{1}L^{\,}_{2}}
\sum_{\bs{q},\bs{k}}
\sum_{s} 
e^{\mathrm{i}\,\bs{q}\cdot\bs{r}}\, 
c^{\dag}_{\bs{k}+\bs{q},s}\,
c^{\   }_{\bs{k},s},
\end{equation}
and construct the matrix with elements
\begin{equation}
{\varrho}_{\bs{r};ij}:=
\langle\Psi^{\,}_{i}|\,
\rho^{\,}_{\bs{r}}\,
|\Psi^{\,}_{j}\rangle.
\label{eq: rhomatrix}
\end{equation}
\end{subequations}

A set of $n$ maps of the local fermion density in the
ground-state manifold is obtained as follows for some finite
lattice $\Lambda$. 
We denote with 
$v^{(i)}_{\rho;\bs{r}^{\,}_{0}},\ i=1,\cdots,n$ 
the set of orthonormal eigenvectors of 
${\varrho}^{\,}_{\bs{r}^{\,}_{0}}$
at some arbitrarily chosen site $\bs{r}^{\,}_{0}$
and evaluate the $n$ real functions
\begin{equation}
n^{(i)}_{\bs{r}}:=
v^{(i)\dagger}_{\rho;\bs{r}^{\,}_{0}}
\;\,
{\varrho}^{\,}_{\bs{r}}\;\,
v^{(i)}_{\rho;\bs{r}^{\,}_{0}}
\;,
\qquad i=1,\cdots,n.
\label{eq: density n(r)}
\end{equation}
The functions $n^{(i)}_{\bs{r}}$ are density maps of the $n$ 
linearly independent combinations of the states 
$|\Psi^{\ }_{1}\rangle,\cdots,|\Psi^{\,}_{n}\rangle$ 
selected by the set of eigenvectors 
$v^{(i)}_{\rho;\bs{r}^{\,}_{0}},\ i=1,\cdots,n$.  
These functions show the variation of the local 
fermion density in position space, i.e., 
the finite lattice $\Lambda$. The extrapolation of these
density maps to the thermodynamic limit can be used to 
distinguish between a ground-state manifold that supports 
a CDW, a phase separation in position space, 
or is featureless as would be expected from a FCI. 
We will give two examples. Example 1 is the case
of the cross-over of a fractional Chern state from a 
featureless liquid to a CDW as the aspect ratio of the lattice is
varied. Example 2 is the case of a phase-separated ground state
obtained with attractive interactions between 
the single-particle states from the flat band.

\begin{figure}
\includegraphics[angle=0,scale=0.45]{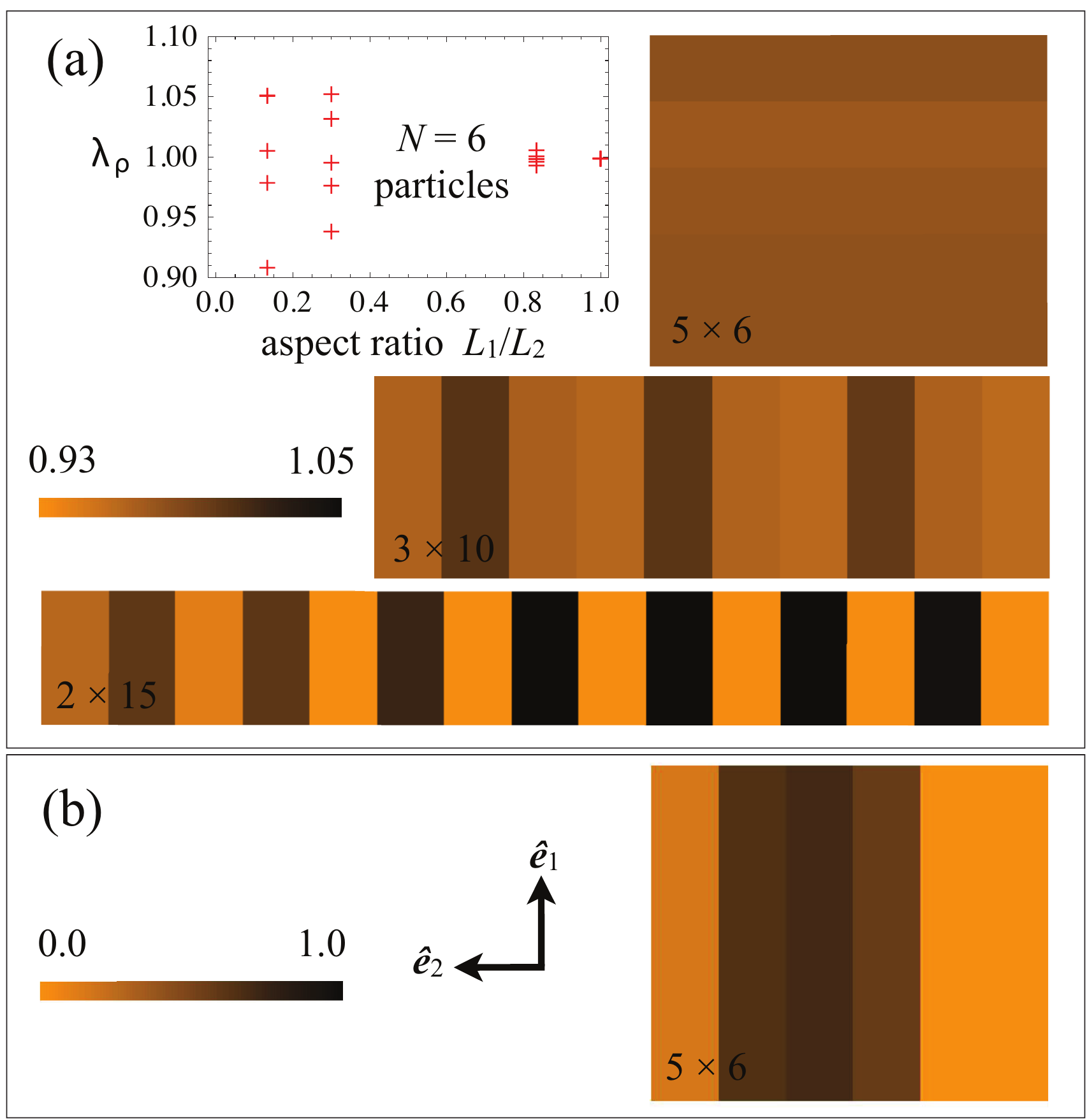}
\caption{
(Color online)
Density profile $n^{(i)}_{\bs{r}}$ 
defined in Eq.~\eqref{eq: density n(r)}
for one representative many-body ground state
$|\Psi^{\,}_{i}\rangle$ 
among the $n$ linearly independent quasidegenerate
many-body ground states.
Note the difference in the color scale between the plots 
in (a) and (b).
(a) Case of the filling fraction $\nu=1/5$ 
when the representative ground states for 
$5\times6$, $3\times10$, and $2\times15$ lattices
would turn into a fractional Chern insulator in 
the thermodynamic limit, by which the two linear dimensions 
of the lattice are much larger
than the correlation length. In a finite lattice,
a CDW profile becomes more pronounced 
when the aspect ratio mimics the thin-torus limit, 
by which the thermodynamic limit is taken with
one of the two linear dimensions of the lattice comparable to or
smaller than the correlation length.
The inset shows the evolution of the eigenvalues $\lambda^{\,}_{\rho}$
of the matrix $\varrho^{\,}_{\bs{r};ij}$ defined in 
Eq.~\eqref{eq: rhomatrix}
as the aspect ratio of the lattice is changed.
(b) Case of the filling fraction $\nu=1/5$ 
when the representative ground state for a $5\times6$ lattice 
would turn into a phase-separated ground state in the thermodynamic
limit, by which the two linear dimensions of the lattice are much larger
than the correlation length, as a result of
attractive nearest-neighbor interactions.
        }
\label{fig: DensityPlots}
\end{figure}

Fractional quantum Hall states on the torus turn smoothly
into a CDW state, if the ratio 
$\min(L^{\,}_{1},L^{\,}_{2})/\ell\lesssim1$.%
~\cite{Tao83,Rezayi94,Bergholz05,Bergholz08} 
Here, $\ell$ is the magnetic length.
The counterpart to this so-called thin-torus limit 
also exists for fractional Chern insulators.%
~\cite{Bernevig12c} 
For concreteness, let us consider the $\nu=1/5$ 
fractional Chern state at Chern number 2.
For the aspect ratio $L^{\,}_{1}/L^{\,}_{2}=1$, 
as is the case for a lattice of $5\times 5$ sites with 
five particles, all five topological ground states 
have the same center-of-mass momentum 
$\bs{Q}=\bs{0}$.
Therefore, 
${\varrho}(\bs{r})$ 
is the unit matrix for all $\bs{r}\in\Lambda$ 
and all functions 
$n^{(i)}_{\bs{r}}$, $i=1,\cdots,5$, are independent of $\bs{r}$.
As a consequence, the fractional Chern state is indeed 
featureless in this isotropic case.
Upon choosing the slightly anisotropic lattice 
$L^{\,}_{1}=5$, $L^{\,}_{2}=6$ with $N=6$ particles,
the eigenvalues of 
${\varrho}(\bs{r})$
lie between $0.994$ and $1.007$.
This gives rise to a small density variation in position space 
of about $1\%$ 
[see Fig.~\ref{fig: DensityPlots}(a)].
Increasing further the anisotropy to 
$L^{\,}_{1}=2$, $L^{\,}_{2}=15$ with $N=6$ particles 
results in a spread of the eigenvalues of 
${\varrho}(\bs{r})$
between $0.91$ and $1.05$
[see Fig.~\ref{fig: DensityPlots}(a)].
In this case, the functions 
$n^{(i)}_{\bs{r}}, i=1,\cdots,5$ 
form the expected CDW pattern with 
pronounced density minima and maxima along the 
$\hat{\bs{e}}^{\,}_{1}$ direction, 
as shown in Fig.~\ref{fig: DensityPlots}(a).

\begin{figure}
\includegraphics[angle=0,scale=0.22]{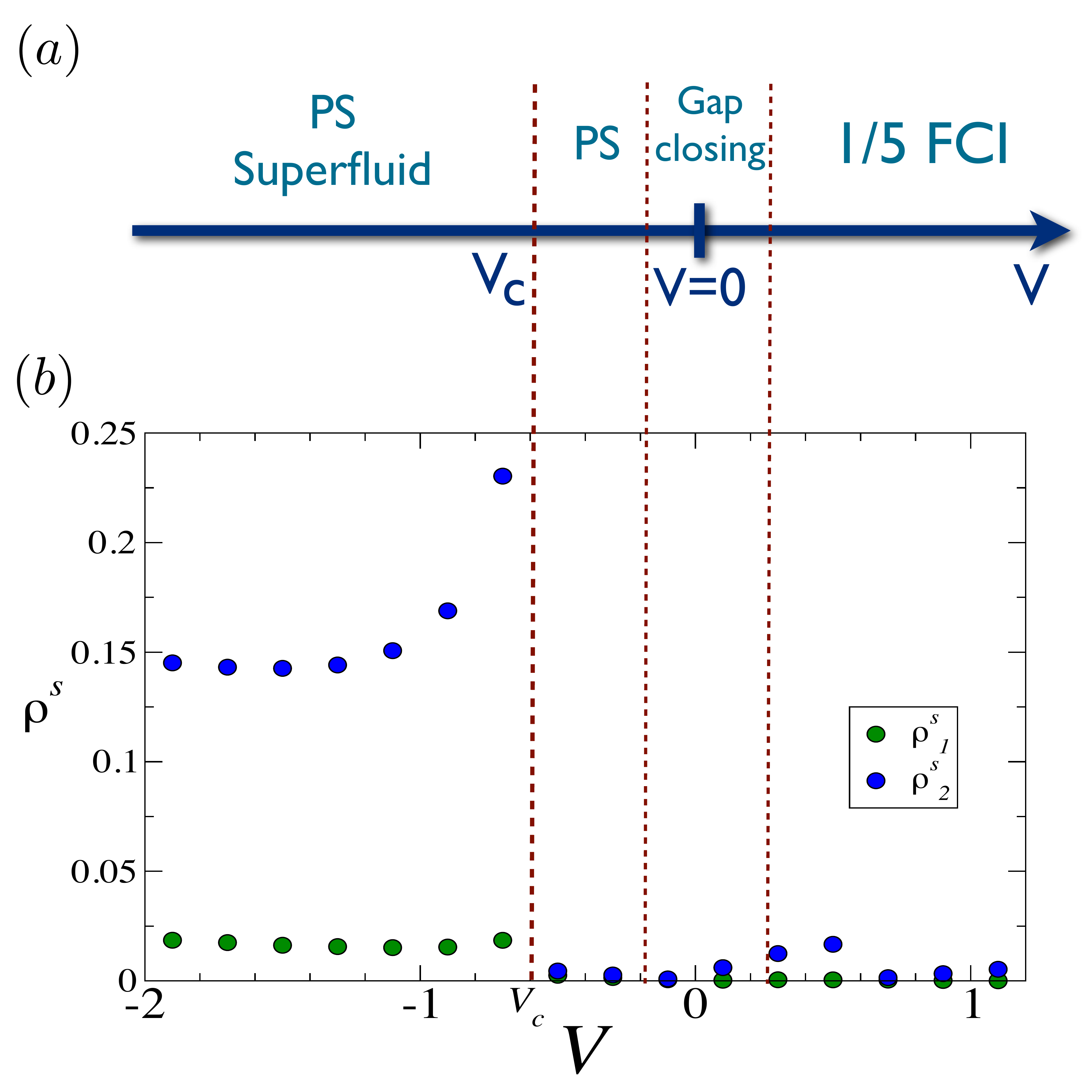}
\caption{\label{fig: PD}
(Color online)
(a)
Quantum phase diagram of
Hamiltonian~(\ref{eq: def Hamiltonian studied by exact dia})
at $\lambda=0$ for fixed noninteracting parameters 
corresponding to two flat bands with $C=\pm2$ 
as a function of the nearest-neighbor interaction strength
$V$ for fixed on-site interaction strength $U=t$. 
The noninteracting parameters are $h^{\,}_{4}=0.7\,t$ 
and $h^{\,}_{i}=0$ 
for $i=1,2,3$. There are three phases. 
For $V>0$ the $\nu=1/5$ fractional Chern insulator (FCI) 
is the most stable phase. 
For $V<0$ the system phase separates (PS) 
and turns superconducting
below a critical $V^{\,}_{\mathrm{c}}\sim -0.6\,t$.
(b)
Superfluid phase stiffness as defined in 
Eq.~\eqref{eq: superconducting condensate fraction} 
for the two orthogonal spanning directions of the lattice.
        }
\end{figure}

We want to contrast these results 
with the density profile of the ground state 
that emerges when attractive instead of repulsive interactions 
are added to the noninteracting flat band model. 
We choose $U=t$ and vary $V$ from positive to 
negative values for a $L^{\,}_{1}=5$, $L^{\,}_{2}=6$ 
with $N=6$ particles. In doing so, 
we encounter two phase transitions towards 
different gapped ground states at about $V\sim 0$ and $V=-0.6\,t$ 
[see Fig. \ref{fig: PD}(b)].
 
First, we obtain a state with macroscopic phase separation 
of the fermions and quasidegeneracy $L^{\,}_{2}=6$. 
The strongly varying density profile in real space 
[see Fig. \ref{fig: DensityPlots}(b)] 
reveals that all particles cluster in a single stripe and 
the degeneracy emerges from shifting this stripe in 
the $\hat{\bs{e}}^{\,}_{2}$ direction across the lattice. 

Second, a state with quasidegeneracy 
$30=L^{\,}_{1}\times L^{\,}_{2}$ 
is obtained with a density profile that shows a stripe 
similar to the first phase. 

To characterize the physical difference between the two phases
stabilized by an attractive $V$, 
we have computed the phase stiffness 
(see Refs.~\onlinecite{Shastry90},
\onlinecite{Sheng03}, and \onlinecite{Melko05}) 
of the ground state $|\Psi\rangle$ 
against twisting the boundary conditions 
away from periodic ones by a complex phase 
$\gamma^{\,}_{i},\ i=1,2$ 
according to Eq.~\eqref{eq: twisted boundary conditions}.
This stiffness equals the condensate fraction 
$\rho^{\mathrm{s}}_{i}$ 
of superfluid pairing in the state $|\Psi\rangle$ and 
it is given by
\begin{equation}
\rho^{\mathrm{s}}_{i}=
\frac{
\partial^2 E^{\,}_{0}(\bs{\gamma})
     }
     {
\partial \gamma^{2}_{i}
     }
\Bigg|^{\,}_{\gamma^{\,}_{i}={0}},
\qquad i=1,2,
\label{eq: superconducting condensate fraction}
\end{equation}
where $E^{\,}_{0}\equiv\langle\Psi|H(\lambda=0)|\Psi\rangle$ 
is the many-body ground-state energy.~\cite{Sheng03}
We find that in the phase with small negative $V$, 
there is no superfluid pairing, 
while the phase at large negative $V$ 
is a one-dimensional superconductor nucleated 
in the stripe of clustered particles. 
The phase stiffness is plotted for both directions in 
Fig.~\ref{fig: PD}(b).
The large increase of 
$\rho^{\mathrm{s}}_{2}$
when $V$ approaches $V^{\,}_{\mathrm{c}}\sim-0.6\,t$
from below suggests a phase transition.
The direction for which the phase stiffness is the largest
and diverging upon approaching the critical value 
$V^{\,}_{\mathrm{c}}\sim-0.6\,t$
from below is
$\hat{\bs{e}}^{\,}_{2}$.
This is the direction of the stripe along which
the electron density is uniform.
Hence, it is plausible to interpret the phase 
$V<V^{\,}_{\mathrm{c}}$
as a stripe-like superconducting phase.

\medskip
\section{The effect of band dispersion}
\label{sec: dispersion}

\begin{figure*}
\includegraphics[angle=0,scale=0.7]{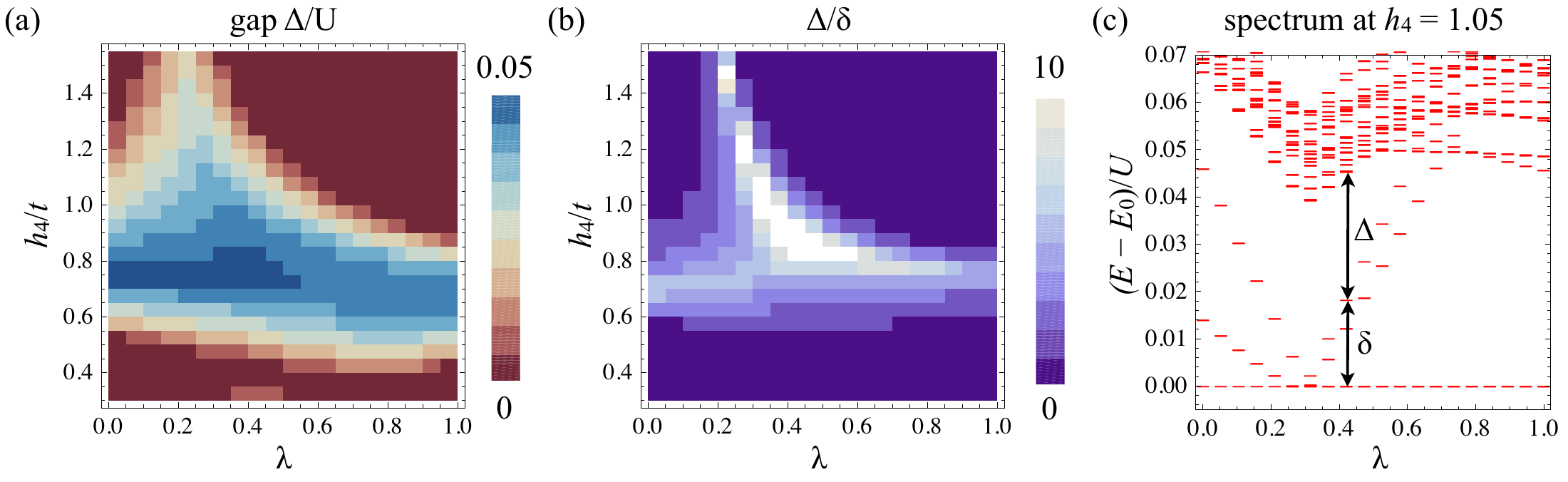}
\caption{\label{fig: finitedisp}
(Color online)
Stability of the FCI phase at $\nu=1/5$ for $C=2$ 
as the bandwidth of the noninteracting Hamiltonian 
is changed via the parameter $\lambda$
defined in Eq.~(\ref{eq: def H0 lambda}).
Panels (a) and (b) show the gap $\Delta$ between 
the fifth- and sixth-lowest 
energy eigenvalue of the many-body Hamiltonian and the
quotient $\Delta/\delta$, where $\delta$ 
is the spread in energy of the five lowest eigenvalues, 
respectively. The region colored blue in (a) and bright (b) 
is interpreted as a FCI with $\nu=1/5$.
In panel (c) the evolution of the lower portion of 
the many-body spectrum with increasing $\lambda$
at constant $h^{\,}_{4}=1.05\,t$ is plotted.
For all figures the number of particles is $N=6$ particles, 
while $L^{\,}_{1}=6$, $L^{\,}_{2}=5$,
$h^{\,}_{i}=0$, $i=1,2,3$, and $U=V=1.5\,t$. 
        }
\end{figure*}

Having established the existence of fractional Chern ground states 
out of band insulators supporting the Chern numbers 1 and 2, 
we now study the effects on FCIs resulting from the spectrum of the
noninteracting Hamiltonian not being flat. 
Bringing back the dispersion of the Bloch bands
amounts to assigning a momentum-dependent energy penalty 
for occupying the single-particle states created 
(by $\chi^{\dag}_{\bs{k}}$) in the BZ. 
Increasing the bandwidth of the noninteracting Bloch bands
to some critical value must result in a phase transition
to a new correlated ground state, 
say a state that supports long-range order 
or a gapless featureless ground state such as
a Fermi liquid. For a Laughlin state at filling $\nu$,
lifting gradually the degeneracy of the Landau levels should
gradually weaken the many-body gap, as we now argue.
After projection into the lowest Landau level, 
the only remaining energy scale is the
(screened) Coulomb interaction that delivers a
many-body correlation length 
$\xi\sim \ell^{\,}_{B}$
[$\ell^{\,}_{B}=\sqrt{\hbar\,c/(eB)}$ the magnetic length].  
Any small one-body perturbation that
breaks Galilean invariance, an impurity potential or
a periodic potential, brings about a 
characteristic length $\ell$. 
The effect of this one-body perturbation on
the distance to a gap-closing phase transition
should depend \textit{solely} on the dimensionless ratio
$\xi/\ell$, in which case a monotonically
decreasing dependence is to be expected.
Indeed, the Berry curvature of a Landau level is constant
in momentum space. It does not favor any particular
finite length scale. Thus, 
the distance to a gap-closing phase transition
upon increasing $\ell$ should depend solely
on the dimensionless ratio $\xi/\ell$.
In contrast, a FCI with flat Bloch bands has a 
Berry curvature that must necessarily vary in
momentum space. The Berry curvature might thus
favor certain characteristic lengths. 
The distance to a gap-closing phase transition
upon increasing $\ell$ should depend on more
than one parameter, in which case non-monotonic
dependence on $\ell$ becomes a possibility.

We use two approaches to support this point. 
First, we give an analytical argument that applies 
to a generic 
lattice model for itinerant interacting fermions in the limit
of a partially occupied flat band.
We show how the addition of a fine-tuned band dispersion 
can be crucial to the selection of a FCI ground state.
Second, we give supporting numerical evidences for 
a scenario by which switching on a finite bandwidth
according to Eq.~(\ref{eq: def H0 lambda}) 
can enhance the stability of a FCI for the specific model 
at hand.

For the analytical argument, 
consider an interacting Hamiltonian of the form
\begin{subequations}
\begin{equation}
H:=
\sum_{\bs{k}^{\,}_{1},\bs{k}^{\,}_{2},\bs{k}^{\,}_{3},\bs{k}^{\,}_{4}}
V^{\,}_{\bs{k}^{\,}_{1}\bs{k}^{\,}_{2}\bs{k}^{\,}_{3}\bs{k}^{\,}_{4}}
\chi^{\dag}_{\bs{k}^{\,}_{1}}
\chi^{\dag}_{\bs{k}^{\,}_{2}}
\chi^{\,}_{\bs{k}^{\,}_{3}}
\chi^{\,}_{\bs{k}^{\,}_{4}}
\end{equation}
that operates exclusively on the Fock space built out of 
the fermion creation operators
$\chi^{\dag}_{\bs{k}},\ \bs{k}\in\mathrm{BZ}$, 
of an isolated flat band. 
This model is generic for translational invariant 
density-density interactions
which have matrix elements of the form
\begin{equation}
V^{\,}_{\bs{k}^{\,}_{1}\bs{k}^{\,}_{2}\bs{k}^{\,}_{3}\bs{k}^{\,}_{4}}=
v^{\,}_{\bs{k}^{\,}_{1}-\bs{k}^{\,}_{3}}
\langle
\chi^{\,}_{\bs{k}^{\,}_{1}}|\chi^{\,}_{\bs{k}^{\,}_{3}}
\rangle
\langle
\chi^{\,}_{\bs{k}^{\,}_{2}}|\chi^{\,}_{\bs{k}^{\,}_{4}}
\rangle
\delta^{\,}_{
\bs{k}^{\,}_{1}+\bs{k}^{\,}_{2},\bs{k}^{\,}_{3}+\bs{k}^{\,}_{4}
            }
\end{equation}
\end{subequations}
in the projected (flat) band,
where $\langle\chi^{\,}_{\bs{k}}|\chi^{\,}_{\bs{k}'}\rangle$
denotes the overlap between a pair of normalized
single-particle Bloch states at $\bs{k},\bs{k}'\in\mathrm{BZ}$
and $v^{\,}_{\bs{k}}$ is the Fourier component
in the BZ of the unprojected and translational invariant 
two-body interaction.

Suppose that the ground state of $H$ at filling $\nu$ is 
a fractional Chern state 
[for instance, this is the case for the Hamiltonian%
~\eqref{eq: H that is diagonalized} with $\nu=1/5$].
We apply a particle-hole transformation in this isolated band. 
This amounts to the transformation
$\chi^{\dag}_{\bs{k}}\to
\chi^{\,}_{-\bs{k}}$,
$\chi^{\,}_{\bs{k}}\to
\chi^{\dag}_{-\bs{k}},\ 
\bs{k}\in\mathrm{BZ}$.
We obtain the transformed Hamiltonian
\begin{subequations}
\begin{equation}
\begin{split}
\widetilde{H}=&\,
\sum_{\bs{k}^{\,}_{1},\bs{k}^{\,}_{2},\bs{k}^{\,}_{3},\bs{k}^{\,}_{4}}
V^{\,}_{
-\bs{k}^{\,}_{1},-\bs{k}^{\,}_{2},-\bs{k}^{\,}_{3},-\bs{k}^{\,}_{4}
       }
\chi^{\dag}_{\bs{k}^{\,}_{3}}\,
\chi^{\dag}_{\bs{k}^{\,}_{4}}\,
\chi^{\,}_{\bs{k}^{\,}_{1}}\,
\chi^{\,}_{\bs{k}^{\,}_{2}}\\
&\,
-\sum_{\bs{k}}
\varepsilon^{\,}_{-\bs{k}}\,
\chi^{\dag}_{\bs{k}}\,
\chi^{\,}_{\bs{k}}
+
\mathrm{constant}
\end{split}
\label{eq: particle-hole transformed H}
\end{equation}
in a normal-ordered form, where
\begin{equation}
\begin{split}
\varepsilon^{\,}_{\bs{k}}
:=&
\sum_{\bs{k}'}
\left(
V^{\,}_{\bs{k}\bs{k}'\bs{k}'\bs{k}}
+
V^{\,}_{\bs{k}'\bs{k}\bs{k}\bs{k}'}
\right)
\\
=&\,
\sum_{\bs{k}'}
\left(v^{\,}_{\bs{k}-\bs{k}'}+v^{\,}_{\bs{k}'-\bs{k}}\right)
|\langle\chi^{\,}_{\bs{k}}|\chi^{\,}_{\bs{k}'}\rangle|^{2}.
\end{split}
\label{eq: def geometric one-body term induced by PH trsf}
\end{equation}
\end{subequations}
We then conclude that, by construction, 
$\widetilde{H}$ supports a fractional Chern state (of holes)
as its ground state at filling $\widetilde{\nu}=1-\nu$
[for instance, $\widetilde{H}$ when derived from
Hamiltonian%
~\eqref{eq: H that is diagonalized} 
at $\nu=1/5$ supports a fractional Chern state (of holes)
as its ground state at the filling $\widetilde{\nu}=4/5$].

This fractional Chern state of holes at 
the filling fraction $\widetilde{\nu}=1-\nu$
is stabilized in the presence of a one-body dispersion
that is of the same order as the interaction itself and
given by $\varepsilon^{\,}_{\bs{k}}$. 
This one-body term can be interpreted as an optimal choice 
of the band dispersion that delivers as the ground state
the fractional Chern state
at the filling fraction $\widetilde{\nu}=1-\nu$.
This one-body term is a trivial constant if and only if 
all overlaps between normalized Bloch states
are functions of $(\bs{k}-\bs{k}')$ only.
Conversely, if overlaps between 
normalized Bloch states also vary as a function of
$(\bs{k}+\bs{k}')$, then the particle-hole transformed
Hamiltonian $\widetilde{H}$ acquires a genuine 
one-body dispersion. Turning off adiabatically
this genuine dispersion can induce a phase transition 
to a ground state that does not display a fractional (hole) 
state at the filling fraction $\widetilde{\nu}$.
For instance, at the filling fraction $\widetilde{\nu}=4/5$,
the Hamiltonians studied numerically in
Refs.~\onlinecite{Laeuchli12} 
and~\onlinecite{Liu12b}
can be interpreted as the Hamiltonian obtained from
$\widetilde{H}$ 
upon subtracting the one-body term $\varepsilon^{\,}_{\bs{k}}$. 
These Hamiltonians at this filling fraction
do not support a fractional Chern ground state,
although $\widetilde{H}$ does.

We close this discussion by observing that
the one-body dispersion%
~(\ref{eq: def geometric one-body term induced by PH trsf}) 
induced by a particle-hole transformation
has an elegant geometric interpretation
that makes it possible to characterize FCIs
though a local quantum metric tensor.
The qualifier quantum originates from the fact that this local 
metric tensor is related to the overlaps of normalized  Bloch states. 
From this geometrical point of view, we are going to show
that the FQHE can be thought of 
as a FCI with a locally flat quantum metric tensor. To carry out this
program, we observe that the one-body dispersion%
~(\ref{eq: def geometric one-body term induced by PH trsf}) 
depends functionally on any one of the quantum distances (metrics) 
\begin{subequations}
\begin{equation}
d^{\,}_{\bs{k},\bs{k}'}(\kappa):=
\sqrt{
1
-
|\langle\chi^{\,}_{\bs{k}}|\chi^{\,}_{\bs{k}'}\rangle|^{\kappa}
     },
\label{eq: def metrics labeled by kappa}
\end{equation}
labeled by the real-valued parameter $\kappa\geq1$ 
between the normalized single-particle Bloch states 
at $\bs{k},\bs{k}'\in\mathrm{BZ}$.
With this definition,
$d^{\,}_{\bs{k},\bs{k}'}(\kappa)$ does indeed satisfy
$d^{\,}_{\bs{k},\bs{k}'}(\kappa)=0$ 
if and only if
$\bs{k}=\bs{k}'$,
$d^{\,}_{\bs{k},\bs{k}'}(\kappa)=d^{\,}_{\bs{k}',\bs{k}}$,
and the triangle inequality 
$d^{\,}_{\bs{k},\bs{k}'}(\kappa)\leq
 d^{\,}_{\bs{k},\bs{k}''}+d^{\,}_{\bs{k}'',\bs{k}'}$
for any triplet of momenta from the BZ.
Now the overlap between any pair of normalized Bloch state 
can be expressed in terms of any one of these metrics,
\begin{equation}
\varepsilon^{\,}_{\bs{k}}=
\sum_{\bs{k}'}
\left(v^{\,}_{\bs{k}-\bs{k}'}+v^{\,}_{\bs{k}'-\bs{k}}\right)
\left[
1
-
d^{2}_{\bs{k},\bs{k}'}(\kappa)
\right]^{2/\kappa}.
\label{eq: one body term and metric}
\end{equation} 
\end{subequations}
The so-called Fubini-Study metric,%
~\cite{Page87}
defined by selecting $\kappa=1$ in 
Eqs.~(\ref{eq: def metrics labeled by kappa})
and
(\ref{eq: one body term and metric}),
plays a special role, for it also enters in 
the algebra of projected density operators as shown 
by Roy in Ref.~\onlinecite{Roy12}. 
For this reason, we select the Fubini-Study metric 
and drop the reference to $\kappa=1$ from now on.
The usefulness of Eq.~(\ref{eq: one body term and metric})
is rooted in the observation that,
in the thermodynamic limit,
if we introduce the local (Fubini-Study) metric tensor 
$g^{\,}_{\mu\nu}(\bs{k})$
through the line integral along the paths $\gamma^{\,}_{1,2}$
connecting the pair $\bs{k}^{\,}_{1}$  and $\bs{k}^{\,}_{2}$ 
\begin{equation}
d(\bs{k}^{\,}_{1},\bs{k}^{\,}_{2})=
\mathrm{inf}^{\,}_{\gamma^{\,}_{1,2}}
\int\limits_{\gamma^{\,}_{1,2}}
\mathrm{d}\ell\,
\sqrt{
g^{\,}_{\mu\nu}(\bs{k})\,
\frac{\mathrm{d}k^{\mu}}{\mathrm{d}\ell}
\frac{\mathrm{d}k^{\nu}}{\mathrm{d}\ell}
     },
\end{equation}
then the local flatness condition
$g^{\,}_{\mu\nu}(\bs{k})\propto\delta^{\,}_{\mu\nu}$
implies that $d(\bs{k}^{\,}_{1},\bs{k}^{\,}_{2})$
is a function of $|\bs{k}^{\,}_{1}-\bs{k}^{\,}_{2}|$ only.
The condition of local flatness on the Fubini-Study metric tensor
has two important consequences. First, the one-body dispersion
generated by the particle-hole transformation on the 
translational invariant two-body interaction is a constant
according to Eq.~(\ref{eq: one body term and metric}), 
as is the case for the Landau levels in the FQHE. 
Second, as shown by Roy in Ref.~\onlinecite{Roy12},
the algebra of projected density operators closes,
as is the case for the Landau levels in the FQHE. Thus, 
it is the departure from a flat (Fubini-Study) metric tensor
that can endow the stability of a FCI with a subtle dependence
on a nonvanishing band-width, as we exemplify next.

We now return to the numerical study of the model defined in 
Eq.~(\ref{eq: def H0 lambda}),
in order to explore the effect of a band dispersion quantitatively. 
To do so,
we need a measure for the stability of a phase that we identify 
by a set of quasidegenerate ground states of small lattices.

One measure for the stability of the phase is simply the size 
of the gap $\Delta$ above the ground-state manifold, i.e., 
the difference in energy between the 
highest of a tuple of quasidegenerate ground-state energies 
and the lowest-energy eigenvalue above it.

A second measure for the stability of the phase is 
the correlation length $\xi$
characterizing the exponential decay of 
the ground-state expectation values of
products of a pair of local operators separated by the distance 
$\bs{r}$. The shorter $\xi$ is, 
the farther is a ground state from a quantum phase transition.
If a manifold of ground states consists of 
topologically degenerate states as is the case for the FCI, 
they become exactly degenerate in the thermodynamic limit
with the property that no local operator can transform one 
of the ground states into another.
For a finite system, in contrast, 
a splitting $\delta$ in energy between the highest
and the lowest of the quasidegenerate ground states
is to be generically expected.
This splitting can be used
as a measure of $\xi$ through the ansatz
\begin{equation}
\frac\Delta\delta\propto e^{L/\xi},
\end{equation}
where $L$ is the characteristic linear size of the system.

For concreteness, we focus on the FCI 
at $\nu=1/5$ with a fivefold quasidegenerate ground state.
We interpolate between the flat band and the original 
noninteracting Hamiltonian with the help of
the family of noninteracting
Hamiltonians~(\ref{eq: def H0 lambda})
parametrized by $\lambda\in[0,1]$ and calculate how the
stability of the candidate FCI phase changes, as measured by
$\Delta$ and $\Delta/\delta$.
Varying $\lambda$ from 0 to 1 
makes it possible to change the bandwidth $W$ 
of the lower band relative to the energy scale of the interaction. 
The bandwidth of the Bloch band is thus given by  
$W=\lambda\,W^{\,}_{0}$, 
where $W^{\,}_{0}$ is the bandwidth of the noninteracting band, 
that depends on the parameters of the model, 
$h^{\,}_{\mu}$ in our case.
We do, however, project the Hilbert space to the one 
spanned by the single-particle states of the lower band for all
values of $\lambda$.

We present in Fig.~\ref{fig: finitedisp} 
both $\Delta$ and $\Delta/\delta$ 
as a function of the bandwidth-parameter $\lambda$
and the parameter $h^{\,}_{4}$ of the noninteracting Hamiltonian%
~\eqref{eq: noninteracting H momentum space}, 
where the bandwidth is 
$W=\lambda\,\mathrm{max}\{\sqrt{2},|\sqrt{2}-h^{\,}_{4}|\}$.
While the largest absolute gap $\Delta$ is indeed obtained 
for the limit of flat bands $\lambda=0$ in this parameter space, 
we observe that for a large range of values of $h^{\,}_{4}$ 
both the gap $\Delta$ as well as the quotient $\Delta/\delta$ can 
be substantially increased as $\lambda$ becomes larger.
We interpret this as an increase in the stability of the FCI phase 
with increasing bandwidth of the noninteracting band and  
exemplify this scenario for $h^{\,}_{4}=1.05\,t$ in 
Fig.~\ref{fig: finitedisp}(c).
Furthermore, we observe that the FCI is stable
against a substantial bandwidth of the order of 
the interaction energy scale for $h^{\,}_{4}\approx 0.7\,t$.

\medskip
\section{
Conclusions
        }
\label{sec: conc}

The search of materials that realize FCIs requires 
the hierarchy of energy scales 
\begin{equation}
W\ll V\ll m,
\end{equation}
with the largest and smallest energy
scale the band gap $m$ and the bandwidth $W$
of the noninteracting band structure, respectively,
while the intermediary energy scale $V$ is the 
characteristic interaction energy. This work 
assumes the limit 
\begin{equation}
W/m\ll V/m\to0
\end{equation}
and deals with the energetic question of what 
is the optimum ratio
of $W/V$ needed to stabilize FCIs.

In this work we have established that 
fractional Chern insulator
phases can be stabilized in situations that differ 
considerably from the ordinary Landau level paradigm, 
namely a flat band-with Chern number 1. 
We have characterized the phase diagram of a model that
hosts Bloch bands with Chern number either one or two 
by means of exact diagonalization. Together with the usual 
characterizations via flux insertion and the counting of 
degenerate ground states, we have mapped out the real space 
density profile of several phases that appear in
this model at filling $\nu=1/5$ for $C=2$: 
CDW, phase separation, 
and fractional Chern insulator. We showed that the gaps
for the $\nu=1/5$ fractional Chern insulating state
with $C=2$ are more than one order of magnitude larger 
than those for the most stable state 
that we found in the model for $C=1$, at $\nu=1/3$. From
these studies we conclude that the most favorable conditions 
to stabilize a FCI need not be those with $C=1$.

By analyzing the phase boundaries of the $\nu=1/5$ 
fractional Chern
insulating state for $C=2$, we have found instances 
for which the flat band condition is not optimal. 
Instead a nonvanishing bandwidth is. 
In these examples, the stability of the FCI first
increases with bandwidth, reaches a maximum, and then 
decreases again as too large a bandwidth compared to 
the interaction energy scale
disfavors the topological state. We have thus shown that 
it is possible to enhance the stability of a
fractional Chern insulator by moving away 
from the condition of flat bands.

The energetic question of when electron-electron
interactions can select strongly correlated phases of matter 
supporting topological order is much
more subtle when electrons populate the Bloch levels
of a Chern band insulator instead of the lowest Landau level. 
This difference manifests itself
with the fact that the Berry curvature is never constant
throughout the BZ in the former case,
whereas it is perfectly flat for the lowest Landau level 
in the latter case.
In this sense, the competition between the kinetic energy
and the electron-electron interaction is much richer for
FCIs than it is for the FQHE from the lowest Landau level.
These subtleties highlight the importance of the interplay
between topology and energetics.

\medskip
\section*{Acknowledgments}

We thank A.\ Bernevig and R.\ Thomale for 
insightful discussions.
This work was supported in part by DOE Grant DEFG02-06ER46316
and by the Swiss National Science Foundation. A.\ G.\ G.\ 
acknowledges 
conversations with M.\ A.\ H.\ Vozmediano and support from 
Spanish grants
FIS2008-00124, FIS2011-23713 and PIB2010BZ-00512 (Brazil).


\begin{thebibliography}{99}
\bibitem{footenote: kappa(i) does not depend on r} 
The transformation law obeyed by 
${\cal O}(\bs{r})$ 
under the translation
$\bs{r}=\bs{r}'+(\bs{r}-\bs{r'})$
is
$
\mathcal{O}(\bs{r})=
U^{\dag}_{\bs{r}-\bs{r'}}\;
\mathcal{O}(\bs{r'})\,
U^{\,}_{\bs{r}-\bs{r'}},
$
with the unitary matrix
$
U^{\,}_{\bs{r}}:=
\mathrm{diag}(
e^{\mathrm{i}\,\bs{r}\cdot\bs{Q}^{\,}_{1}},
\cdots,
e^{\mathrm{i}\,\bs{r}\cdot\bs{Q}^{\,}_n})
$
that represents the action of the translation operator
in the basis that we chose for the ground-state manifold.
\end{thebibliography}
\end{document}